\documentclass[pra,twocolumn,floatfix,showpacs]{revtex4-1}
\usepackage{graphicx,color}
\usepackage{amsmath,amssymb,bm}

\newcommand{\tn}{\textnormal}

\newcommand{\cprb}[3]{Phys.~Rev.~B {\bf #1}, #2 (#3)}
\newcommand{\cprl}[3]{Phys.~Rev.~Lett.~{\bf #1}, #2 (#3)}

\definecolor{darkred}{rgb}{0.90,0,0}
\definecolor{darkgreen}{rgb}{0,0.60,.2}
\definecolor{darkblue}{rgb}{0,0,1}
\definecolor{grey}{cmyk}{0,0,0,0.25}
\definecolor{orange}{cmyk}{0,0.6,0.8,0}

\begin{document}

\title{Entanglement scaling of excited states\\in large one-dimensional many-body localized systems}

\author{D.\ M.\ Kennes$^1$}

\author{C.\ Karrasch$^2$}

\affiliation{$^1$Institut f\"ur Theorie der Statistischen Physik, RWTH Aachen University and JARA-Fundamentals of Future Information Technology, 52056 Aachen, Germany}
\affiliation{$^2$Dahlem Center for Complex Quantum Systems and Fachbereich Physik, Freie Universit\"at Berlin, 14195 Berlin, Germany}

\begin{abstract}

We study the properties of excited states in one-dimensional many-body localized (MBL) systems using a matrix product state algorithm. First, the method is tested for a large disordered non-interacting system, where for comparison we compute a quasi-exact reference solution via a Monte Carlo sampling of the single-particle levels. Thereafter, we present extensive data obtained for large interacting systems of $L\sim100$ sites and large bond dimensions $\chi\sim1700$, which allows us to quantitatively analyze the scaling behavior of the entanglement $S$ in the system. The MBL phase is characterized by a logarithmic growth $S(L)\sim \log(L)$ over a large scale separating the regimes where volume and area laws hold. We check the validity of the eigenstate thermalization hypothesis. Our results are consistent with the existence of a mobility edge. 

\end{abstract}

\maketitle



\section{Introduction}

Disorder such as impurities or vacancies is present in every physical system. Our basic understanding of its effects goes back to Anderson \cite{anderson}: Roughly speaking, the ratio between the electron wavelength and the mean free path determines whether states are extended or localized and hence whether the system is a metal or an insulator. In one or two spatial dimensions, an arbitrarily small amount of disorder will localize any eigenstate in the spectrum, but in 3d a so-called mobility edge can exist which separates localized states at the lower end of the spectrum from extended states at high energies. The transition, which can, e.g., be triggered by varying the disorder strength, is the so-called Anderson transition. Three distinct properties of the localized phase are that a) the DC conductivity $\sigma$ vanishes in the thermodynamic limit at any finite temperature (if all single-particle states are localized, i.e.~in absence of a mobility edge), b) the system does not thermalize, but information about an initial state is preserved in local observables during a unitary time evolution, and c) the entanglement entropy is not extensive but features an area law. Reviews of these single-particle localization physics can be found in Refs.~\cite{andrev1,andrev2,andrev3}.

Anderson localization is formulated for free particles; one might wonder whether interactions delocalize every state, render the conductivity finite and lead to thermalization. In a seminal work, however, Basko, Aleiner, and Altshuler suggested that the localized phase can exist even in presence of interactions and that a finite temperature phase transition can exist between phases with $\sigma=0$ and $\sigma>0$ \cite{mbl0}. This phase transition is not a thermodynamic (equilibrium) transition but a dynamical quantum phase transition which occurs on the level of the many-body eigenstates and defies standard Mermin-Wagner arguments.

In contrast to localization in a non-interacting system, the world of many-body localization (MBL; see Ref.~\cite{mblrev} for a recent review) is still comparably young. Basko, Aleiner and Altshuler's idea was based on perturbative arguments whose range of validity for a given microscopic model is a priori unclear. For one-dimensional lattice systems, the stability of localized states towards adding interactions -- i.e., the existence of the MBL phase -- has subsequently been established fairly convincingly by a number of numerical \cite{mblhuse1,mblprosen,mbled1,mblhuse2,mblarea,mbljens2,mblvasseur,edluitz} and analytical \cite{mbltop1,mbltop2,mblproof} studies. Moreover, there is now solid evidence that a transition into a delocalized phase occurs if the ratio between the interaction and the disorder strength is increased \cite{mblhuse2,mbljens2,mblvasseur,edluitz,mblclusterexp}. The MBL phase was characterized via level statistics \cite{mblhuse1,mblhuse2}, entanglement measures \cite{mblarea,mbljens,mbljens2,mblstates2}, thermalization behavior \cite{mblitaly,mblhuse2,mblscardicchio}, or integrals of motion \cite{mblstates1,mblintmotion}. Some experimentally-accessible observables were computed such as transport properties \cite{mbltrans1,mbltrans2,mbltrans3,mblehud,mblehud2,mbltrans4,mbltrans5,mbltrans6,mbltrans7} or spectral features in presence of a bath \cite{mblbath1,mblbath2}. MBL physics was observed experimentally in a cold atom setup \cite{mblexp,mblexp2}.

However, even in 1d there is a variety of open questions. First, most of the above-mentioned studies were based on an exact diagonalization of small systems, and since it is unclear how to properly perform a finite-size scaling, very little is known about the nature of the dynamical phase transition between the MBL and the metallic phases (e.g., its universality class). Second, the issue of a many-body mobility edge in 1d (its existence and how it depends on the model parameters) is still debated \cite{mbledge}, and so is the ensuing question of a quantitative description of the MBL transition as a function of the temperature.

One route to deepen our understanding of many-body localization physics is to approach the problem using different methods which have their own strengths and shortcomings. In particular, it would be highly desirable to study systems which are larger than those accessible by exact diagonalization. In one dimension, the density matrix renormalization group (DMRG) allows to elegantly compute ground states (or finite-entanglement approximations thereof) as well as those excited states which correspond to ground states in different symmetry sectors. In order to describe MBL physics, however, one needs access to generic excited states. Even though many-body localized states feature an entanglement area law and can thus in principle be expressed efficiently as a matrix product state (MPS) \cite{mbleisert}, no algorithm exits to determine this MPS representation in practice (see below for comments on three recent preprints).

It is our goal to introduce a very simple MPS-based framework to calculate excited states of MBL systems. We first perform various internal convergence checks and moreover test the method for the non-interacting case, where we construct eigenstates at a given energy via a Monte Carlo sampling of the known single-particle levels. We then present extensive data for large systems of $L\sim100$ sites obtained for large bond dimensions of $\chi\sim1700$. This allows us to quantitatively analyze the scaling of the entanglement entropy. We also verify the violation of the eigenstate thermalization hypothesis. Indications of a crossover into the metallic phase and the existence of a mobility edge are presented.

Shortly before the completion of our work, we became aware of three preprints \cite{mbldmrg1,mbldmrg2,mbldmrg3} which present similar ideas to access excited states of MBL systems via DMRG algorithms (these algorithms are more elaborate than the one used here); another method to compute features of the entire spectrum rather than individual states was introduced in preprint \cite{mblall}. We complement these studies in the following way: (a) Our data was obtained over the course of 9 months using large-scale numerics (900.000 core hours); this allows us to access systems which are larger than those investigated in Refs.~\cite{mbldmrg1,mbldmrg2} and to employ bond dimensions $\chi\sim1700$ which are much higher than those used in Refs.~\cite{mbldmrg1,mbldmrg2,mbldmrg3}, which in turn yields new quantitative insights about the precise scaling of the entanglement in the MBL phase. (b) We illustrate how in the non-interacting case an exact solution can be constructed using Monte Carlo sampling; since this limit is non-trivial for the DMRG, it provides a non-trivial testing ground for any future algorithmic improvements. (c) We re-examine the question of a mobility edge from an entanglement perspective.

\section{Model and Method}

\subsection{Model}
\label{sec:model}

We consider one-dimensional spinless interacting fermions living on a lattice of size $L$, or equivalently, a XXZ spin chain:
\begin{equation}\label{eq:h}\begin{split}
H = \sum_{l=1}^{L}\Big(\frac{1}{2}S_l^+S_{l+1}^- + \tn{h.c.} + \Delta S_l^zS_{l+1}^z + V_lS^z_l\Big),
\end{split}\end{equation}
where $S^{x,y,z}$ are spin-1/2 operators, and $S^\pm=S^x\pm i S^y$. The on-site potentials are drawn from a uniform random distribution:
\begin{equation}
V_l \in [-\eta,\eta]\,.
\end{equation}
Prior numerics suggest that a transition between a fully many-body localized and a metallic phase occurs around $\eta\sim3.5$ (see, e.g., Ref.~\cite{edluitz} for an exact diagonalization study of up to $L=22$ sites).

\subsection{DMRG for excited states}
\label{sec:dmrg}

The density matrix renormalization group \cite{white1,dmrgrev} provides an algorithm to variationally compute the ground state within the class of matrix product states. The matrix (bond) dimension $\chi$ encodes the amount of entanglement $S$ in the system. If the problem at hand features an area law, $S$ is non-extensive, and the ground state can thus be represented exactly by a MPS with a finite $\chi$ even in the thermodynamic limit. More generally, one can think of the DMRG as a tool to determine finite-entanglement approximations.

Generalizations of the DMRG algorithm allow to compute (approximations to) a few low-lying excited states as well as those excited states which correspond to ground states in a different symmetry sector \cite{dmrgrev}. However, a practical way to extract arbitrary states in the spectrum does not exist even if it is known that they can in principle be expressed efficiently by a MPS due to their finite entanglement; this is exactly the case in localized systems. Moreover, many-body localization cannot be understood by only considering ground states, and developing a tool to determine the MPS representation of lowly-entangled excited states is thus desirable.

We introduce a simple scheme to calculate finite-entanglement approximations of generic excited states in localized systems. Its basic idea is to consider a set of auxiliary operators $f_\lambda(H)$ whose ground states correspond to excited states of the original Hamiltonian $H$. This implies that any existing DMRG code can be used straightforwardly for the calculation. The only a priori requirement is that $f_\lambda(H)$ can be written as a matrix product operator of low dimension. Here, we employ (we will comment on potential pitfalls associated with this choice below)
\begin{equation}\label{eq:fh}
f_\lambda(H) = \lambda H + (1-\lambda) H^2\,,
\end{equation}
which for the Hamiltonian of Eq.~(\ref{eq:h}) can be written as a MPO of dimension $9$:
\begin{equation}
W^{[l]}=
\begin{pmatrix}
M_1^l & 0 \\ M_2^l & M_3^l
\end{pmatrix}\,,
\end{equation}
where for $l\neq 1, L$:
\begin{equation}
M_1^{l}=
\begin{pmatrix}
I&0&\ldots\\
\frac{S^-}{2}&0&\ldots\\
\frac{S^+}{2}&0&\ldots\\
S^z&0&\ldots\\
\end{pmatrix}\,,~
M_3^{l}=
\begin{pmatrix}
\vdots & \vdots & \vdots & \vdots \\
0 & 0 & 0 & 0 \\
S^+&S^-&S^z&I
\end{pmatrix}\,,
\end{equation}
and
\begin{widetext}
\begin{equation}
M_2^l=
\begin{pmatrix}
2V_{l}(1-\lambda)\Delta&2(1-\lambda)S^+&2(1-\lambda)S^-&2\Delta(1-\lambda)S^z&I\\
0& \frac{1-\lambda}{2}I&0&0&\frac{S^-}{2}\\
0&0&\frac{1-\lambda}{2}I&0&\frac{S^+}{2}\\
0&0&0& \frac{\Delta^2(1-\lambda)}{2}I&\Delta S^z\\
\lambda V_{l}   + \frac{V_{l-1}+V_{l+1}}{2}\Delta(1-\lambda) S^z
&\left[\lambda-\frac{\Delta(1-\lambda)}{2}\right]S^+&\left[\lambda-\frac{\Delta(1-\lambda)}{2}\right]S^-&\left[\Delta\lambda- \frac{1-\lambda}{2}\right]S^z&V_{l}(1-\lambda)S^z
\end{pmatrix}\,.
\end{equation}
\end{widetext}
We determine the ground state of $f_\lambda(H)$ using a standard two-site DMRG algorithm \cite{dmrgrev}. The discarded weight is fixed (we have checked that lowering it further does not change our results); hence, the bond dimension $\chi$ increases during the DMRG sweeps. Our calculations are carried out using even system sizes $L$ and in a fixed symmetry sector $\langle\sum_{l=1}^LS^z_l\rangle=0$. Convergence is checked via the variance of the energy; we allow for a comparably large value $\tn{var}(H)\sim 10^{-6}\gg e^{-L}$. Hence, we can \textit{a priori} only expect to obtain superpositions of nearby eigenstates, which can lead to an artificially-increased entanglement. This issue needs to be investigated carefully (see below).

\begin{figure}[b]
\includegraphics[width=0.95\linewidth,clip]{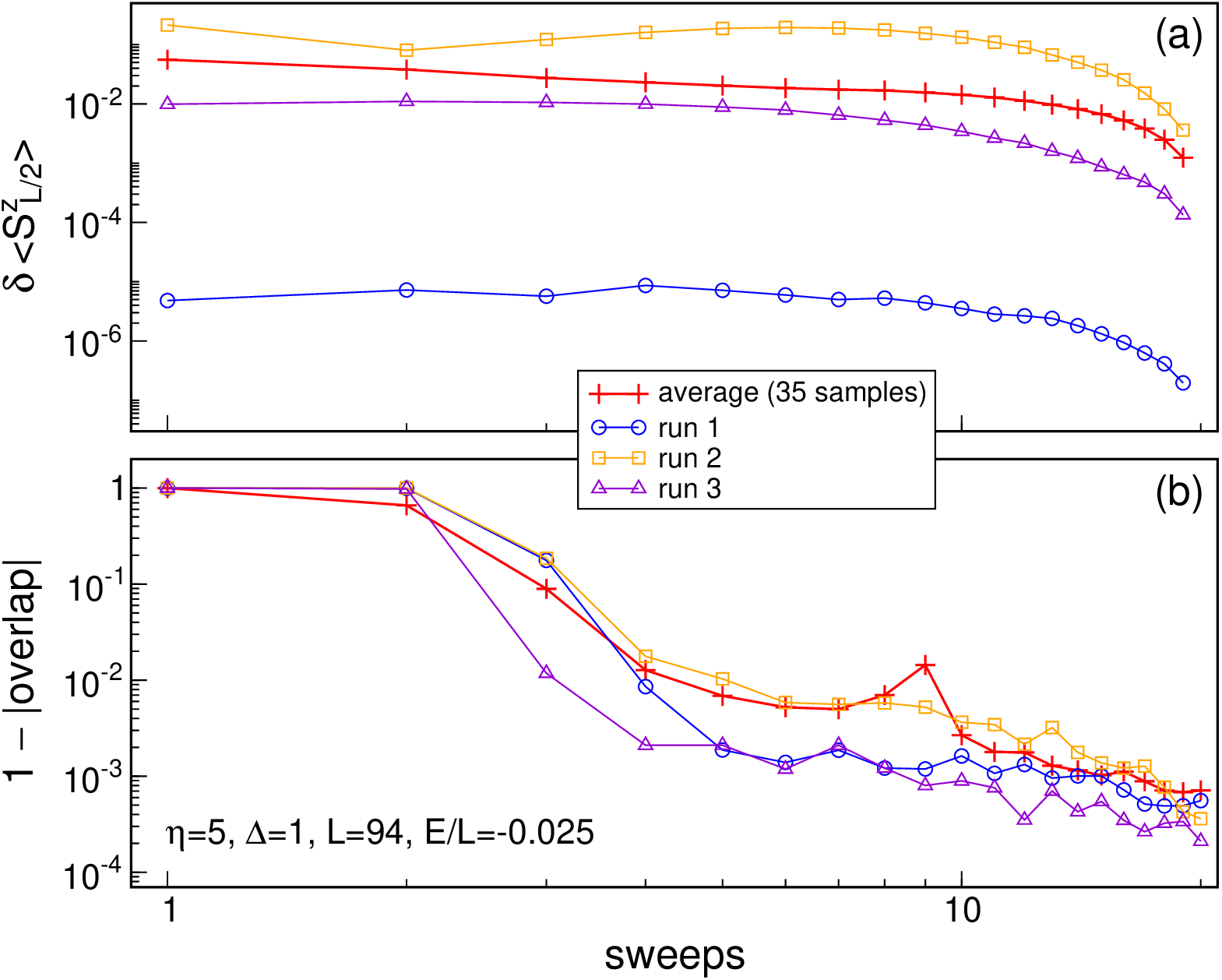}
\caption{(Color online) Convergence tests for the DMRG algorithm in the MBL phase. (a) Magnetization at the center site as a function of the DMRG sweeps (the magnetization reached at the end of the simulation is subtracted). (b) Mutual overlap of the states obtained after two consecutive sweeps. }
\label{fig:szoverlap}
\end{figure}

\begin{figure}[b]
\includegraphics[width=0.95\linewidth,clip]{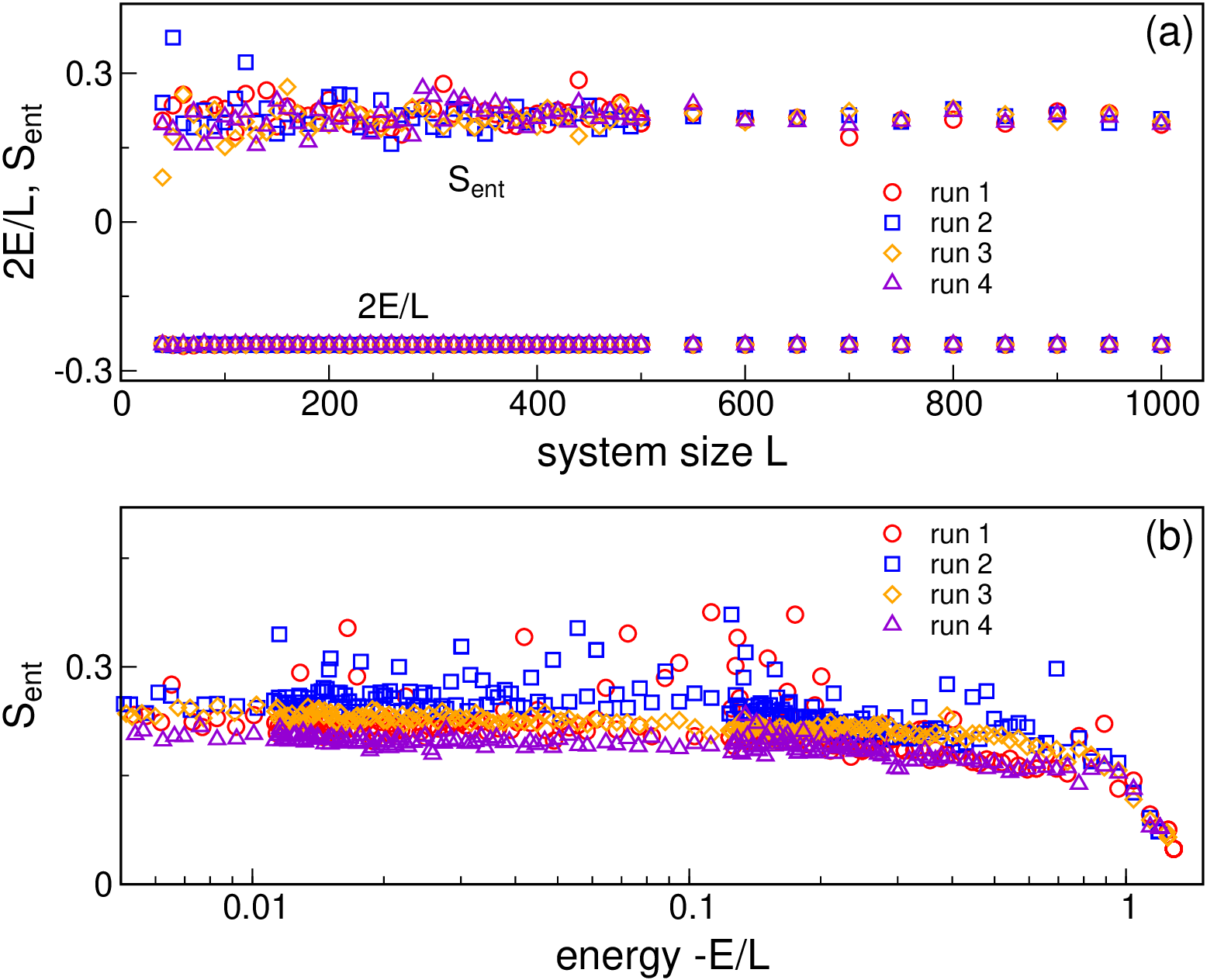}
\caption{(Color online) Proof-of-principle application of the excited-state DMRG algorithm to a non-interacting system ($\Delta=0$) with disorder $\eta=5$. (a) System-size dependence of the entanglement entropy in highly-excited states with an energy density $E/L\approx-0.125$; the area law holds. Four different disorder realizations are shown at each $L$ [$4*(\tn{number of $L$-points})$ realizations in total]. The data was obtained using a mixing $\lambda$ strictly chosen according to the simple form $E=\lambda/[2(\lambda-1)]$. (b) The same but for a constant system size $L=200$ and four different disorder configurations. }
\label{fig:scan}
\end{figure}

In order to gain some further understanding of the capabilities as well as the potential pitfalls of this algorithm, it is instructive to consider the case of free fermions in absence of disorder ($\Delta=\eta=0$). The spectrum of $H$ is then simply characterized by the single-particle states $\epsilon_k=-\cos(k)$, and it is intuitively clear that by varying $\lambda$ one can in principle access excited states at arbitrary energies: While $\lambda=0$ yields the ground state of $H$, the spectrum of $f_{\lambda=1}(H)=H^2$ is governed by $\epsilon_k^2=\cos(k)^2$, and its ground state corresponds to a zero-energy (mid-spectrum) state. However, it is also intuitively clear that this state is \textit{special} in the sense that it is always symmetric with respect to $k=\pi/2$ -- by construction, our algorithm lacks the capability to describe other (asymmetric) zero-energy states. While this specific issue only occurs at $\lambda=1$, it gives rise to the general question of whether or not the states targeted by our method are generic. We will come back to this below.

\begin{figure}[t]
\includegraphics[width=0.95\linewidth,clip]{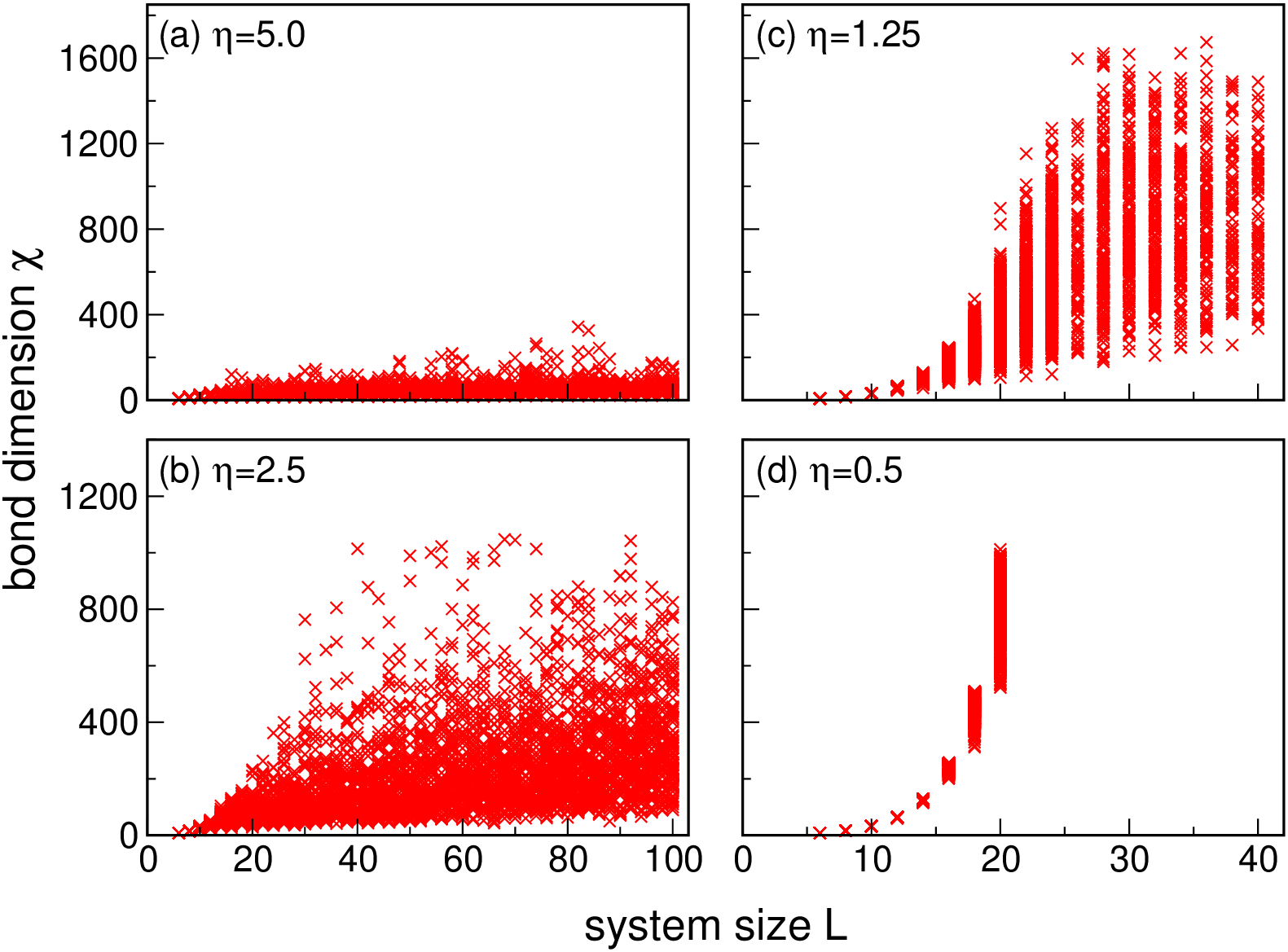}
\caption{(Color online) Scatter plot of the bond dimensions $\chi$ occurring in the DMRG calculation of a highly-excited state with energy $E/L=-0.125$ in a non-interacting system ($\Delta=0$) of size $L$. (a--d) show various disorder strengths $\eta$; the total number of configurations in each panel is $O(5000)$. }
\label{fig:scatter1}
\end{figure}

To summarize, the following potential problems/questions need to be addressed: a) Does our algorithm successively converge to a single eigenstate or to superpositions of eigenstates? b) Are the states/results that one obtains from our simple algorithm generic? c) How do we need to choose $\lambda$ in order to target states at a given energy density?

In order to investigate these questions, it is helpful to solve the non-interacting, disordered problem ($\Delta=0, \eta>0$) analytically; since this limit is not special for the DMRG, it provides an unbiased frame of reference. We will now demonstrate how such an analytic solution can be obtained.

\subsection{Exact solution at $\Delta=0$}
\label{sec:exact}

\begin{figure}[t]
\includegraphics[width=0.95\linewidth,clip]{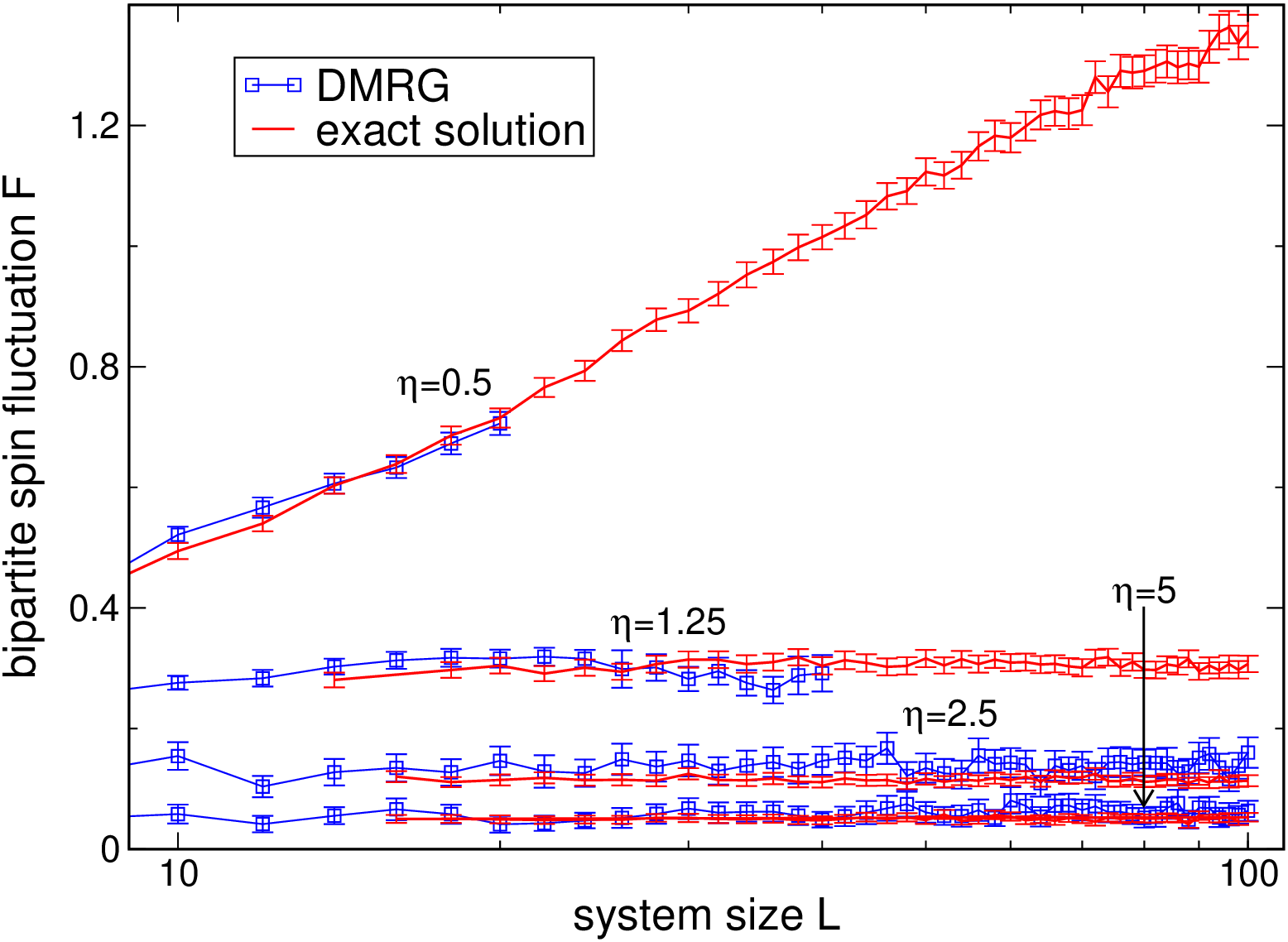}
\caption{(Color online) Scaling of the bipartite spin fluctuations $F(L)$ corresponding to the data sets shown in Fig.~\ref{fig:scatter1}. For comparison, an analytic reference solution is constructed from the exact diagonalization of the non-interacting Hamiltonian combined with a Monte Carlo sampling of the single-particle levels to determine excited states with $E/L=-0.125$. The error bars are defined via the standard deviation of $F$. }
\label{fig:test}
\end{figure}

At $\Delta=0$, the Hamiltonian of Eq.~(\ref{eq:h}) maps to spinless, non-interacting fermions and can thus be solved exactly. For any given disorder realization, the single-particle energies $\epsilon_i$ are simply obtained by diagonalizing a $L\times L$ matrix, and the many-body eigenstates are given by arbitrarily filling up $L/2$ of these levels (which corresponds to zero magnetization in the spin language). In order to obtain an eigenstate at (approximately) a given energy $E/L$, we need to find the occupation numbers $n_i\in\{0,1\}$ for which $\sum_i n_i\epsilon_i\approx E$. Since there are $L!/(L/2!)^2$ possibilities to occupy half of all single-particle levels, this combinatorial problem cannot be solved straightforwardly for $L\sim100$; instead, we propose to employ the following Monte Carlo algorithm:

First, we determine the temperature $T$ of a Fermi distribution such that
\begin{equation}
 \sum_{i=1}^{L/2} f_i(T)\epsilon_i = E\,,~f_i(T) = \frac{1}{\exp(\epsilon_i/T)+1}\,.
\end{equation}
Thereafter, we draw $L$ random numbers $s_i\in[0,1]$ and obtain a `test configuration' $\{n_i\}$ via
\begin{equation}
n_i = \begin{cases} 0 & s_i \leq f_i(T) \\ 1 & \tn{otherwise .} \end{cases}
\end{equation}
The configuration is discarded if $\sum_i n_i \neq L/2$. We repeat this procedure a large number of times and eventually pick the configuration for which $E_0=\sum_in_i\epsilon_i$ approximates the given $E$ best. In practice (e.g., for the data shown in Fig.~\ref{fig:test}), this allows us to find a $E_0$ that deviates from $E$ by at most one percent within a few seconds. The entanglement in this state is then computed using the results of Refs.~\cite{sent1,sent2}.

\begin{figure*}[t]
\includegraphics[width=0.48\linewidth,clip]{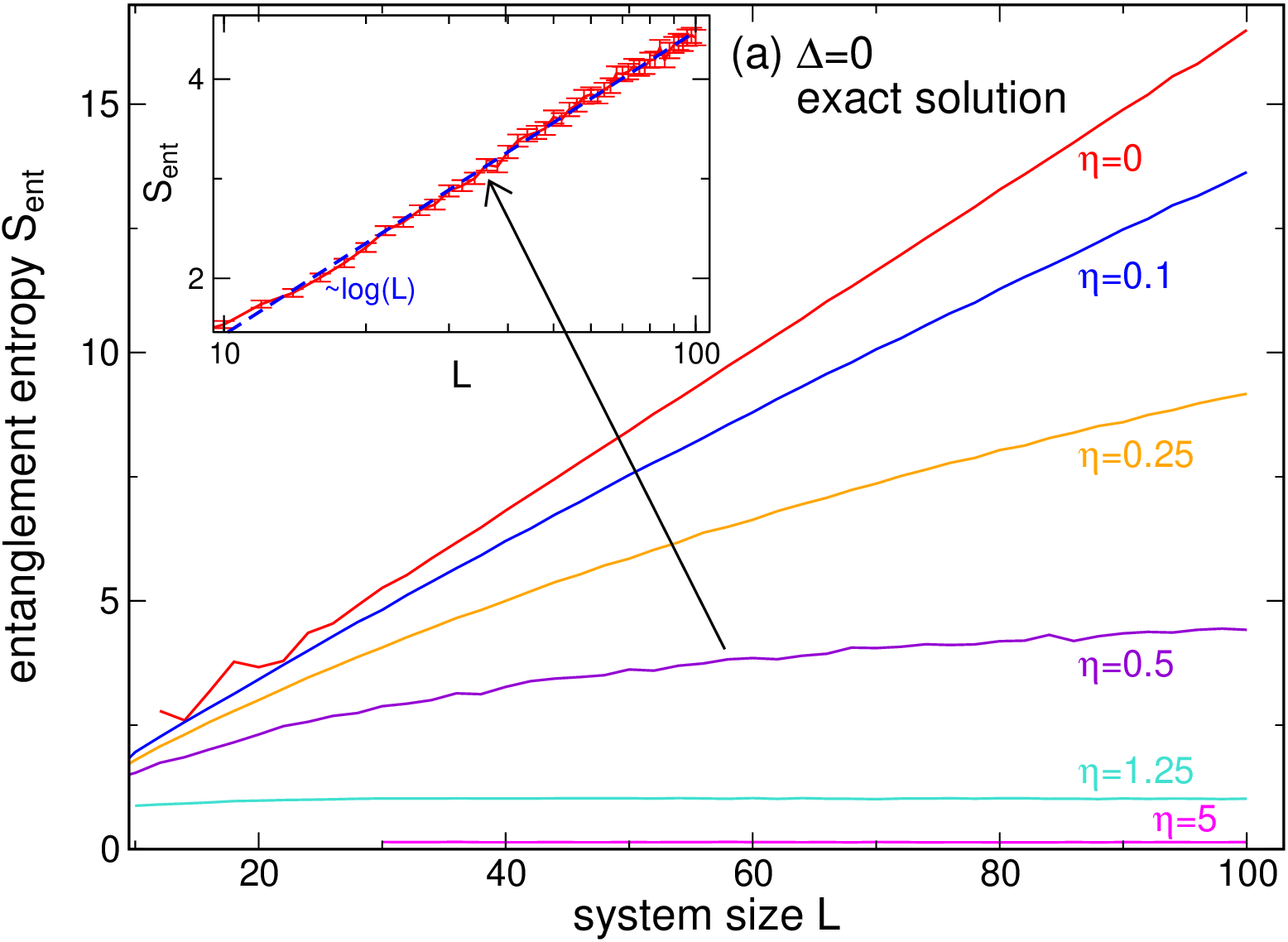}\hspace*{0.02\linewidth}
\includegraphics[width=0.48\linewidth,clip]{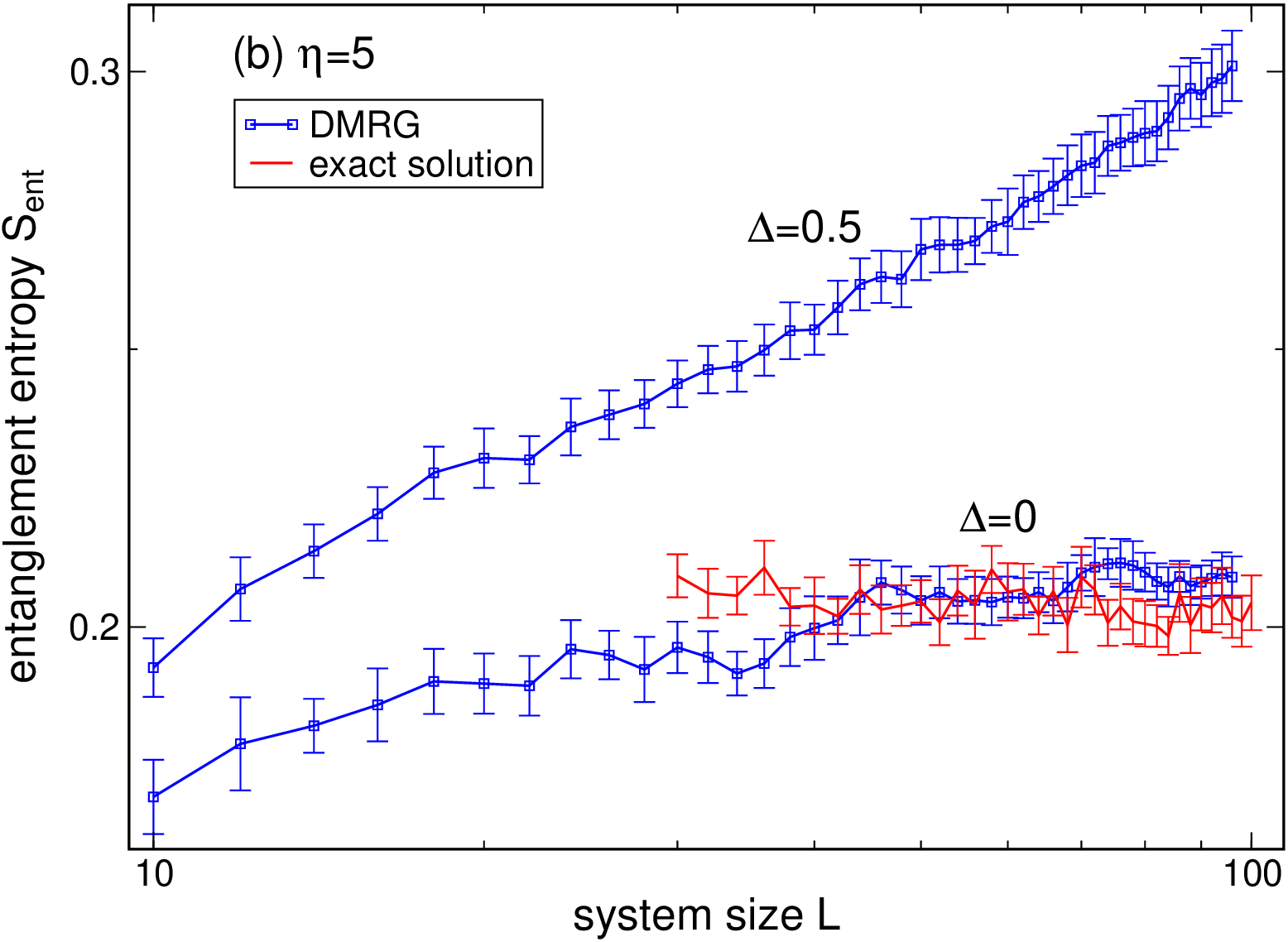}
\caption{(Color online) Scaling of the entanglement entropy in a highly-excited state with $E/L=-0.125$. (a) Exact result for the non-interacting system ($\Delta=0$) and various $\eta$. Upon switching on the disorder, the volume law $S(L)\sim L$ is replaced by a logarithmic increase $S(L)\sim\log(L)$ (see the inset). The area law $S(L)\sim\tn{const.}$ only manifests on a much larger scale [see Fig.~\ref{fig:test3}(a)]. The error is of the order of the linewidth. (b) Comparison of $\Delta=0$ with the interacting case $\Delta=0.5$ at fixed $\eta=5$. In the former case, both the exact solution and DMRG data are shown.}
\label{fig:sent}
\end{figure*}

\section{Tests of the Algorithm}
\label{sec:testresults}

In this section, we subject our algorithm to various internal tests in order to address the issues raised above. This will be complemented by the comparison with the analytic result at $\Delta=0$ presented in the next section.

We first investigate what one can learn about the nature of the states that our algorithm converges to. The (Anderson or many-body) localized phase is characterized by the lack of thermalization. Hence, two eigenstates which are close in energy will in general exhibit vastly different expectation values for local observables (we will explicitly demonstrate this violation of the eigenstate thermalization hypothesis in Sec.~\ref{sec:mblresults.eth}). If our algorithm did successively converge towards some superposition in an uncontrolled way, one would hence expect these expectation values to strongly fluctuate between different DMRG sweeps -- this is the very insight that was exploited in the construction of the algorithm introduced in Ref.~\cite{mbldmrg1}.

In Figure \ref{fig:szoverlap}(a), we show the expectation value of the magnetization $\langle S_{L/2}^z\rangle$ at the center of the chain as a function of the DMRG sweeps. The magnetization reached at the end of the simulation is subtracted, and the parameters are chosen such that the system is in the MBL phase. The curves evolve smoothly, and no vast spatial reordering takes place. Moreover, the overlap of the two states before and after a given sweep almost monotonously approaches unity [see Fig.~\ref{fig:szoverlap}(b)] as the number of sweeps is increased. Both observations are evidence that our very simple algorithm \textit{does not} converge towards a superposition of states which are close in energy but feature different expectation values of local observables.

Finally, we demonstrate that the mixing $\lambda$ can be easily chosen such that states at all energies $E/L$ can be accessed. If $E/L$ lies in a dense part of the spectrum, then Eq.~(\ref{eq:fh}) directly yields the relation $E=\lambda/[2(\lambda-1)]$. Deviations are only expected for small systems or if one tries to target an energy close to the ground state energy. As a consistency check, it is instructive to carry out a calculation using a mixing $\lambda$ strictly chosen according to $E=\lambda/2[(\lambda-1)]$. Results are shown in Figure \ref{fig:scan}(a) for four different disorder configurations. For all data shown in this paper and for $L\geq 10$, the choice $E=\lambda/[2(\lambda-1)]$ is sufficient to obtain a targeted energy density $E/L$ with a relative accuracy of at least one percent.

\section{Test Case: Free Fermions}
\label{sec:free}

\subsection{Raw data}

In order to further explore the capabilities and limitations of the excited-state DMRG algorithm, we extensively study the limit $\Delta=0$ where Eq.~(\ref{eq:h}) maps to non-interacting fermions. We first present the raw DMRG data and discuss how disorder averages can be computed.

In absence of interactions, it is known that an arbitrarily small amount of disorder localizes all states in the spectrum. In Figure \ref{fig:scan}(a), we plot the entanglement entropy $S$ of a highly-excited state with an energy density $E/L\approx-0.125$ for systems of up to $L=1000$ sites. At each value of $L$, four different disorder realizations are drawn from a distribution of strength $\eta=5$. Our results illustrate that the area law $S(L)\sim\tn{const.}$ holds, reflective of the fact that this state is localized.

Next, we compute excited states at a fixed energy density $E/L=-0.125$ for a large number of disorder realizations and system sizes. Figure \ref{fig:scatter1} shows a scatter plot of the bond dimensions $\chi$ necessary to describe these states. In general, smaller $\eta$ require larger values of $\chi$, which is reasonable since one expects the localization length and hence the amount of entanglement to increase with decreasing strength of the disorder. Moreover, Fig.~\ref{fig:scatter1} shows the occurrence of rare states with unusually high entanglement.

Only finite bond dimensions are accessible numerically due to the limitation of computational resources. In this work, we abort each calculation once $\chi$ exceeds a value of $\chi\sim1300$ in general and $\chi\sim1700$ for some exemplary cases. In order to reliably compute averaged quantities, we need to ensure that the states dropped in our calculation are only rare states which do not carry any substantial weight. To this end one can, e.g., successively increase the system size and determine histograms of bond dimensions for each $L$. We eventually discard all data for which substantial weight is shifted above the maximally allowed bond dimension [which is $\chi\sim1700$ for the parameters of Fig.~\ref{fig:scatter1}(c) and $\chi\sim1300$ otherwise].

\begin{figure}[t]
\includegraphics[width=0.95\linewidth,clip]{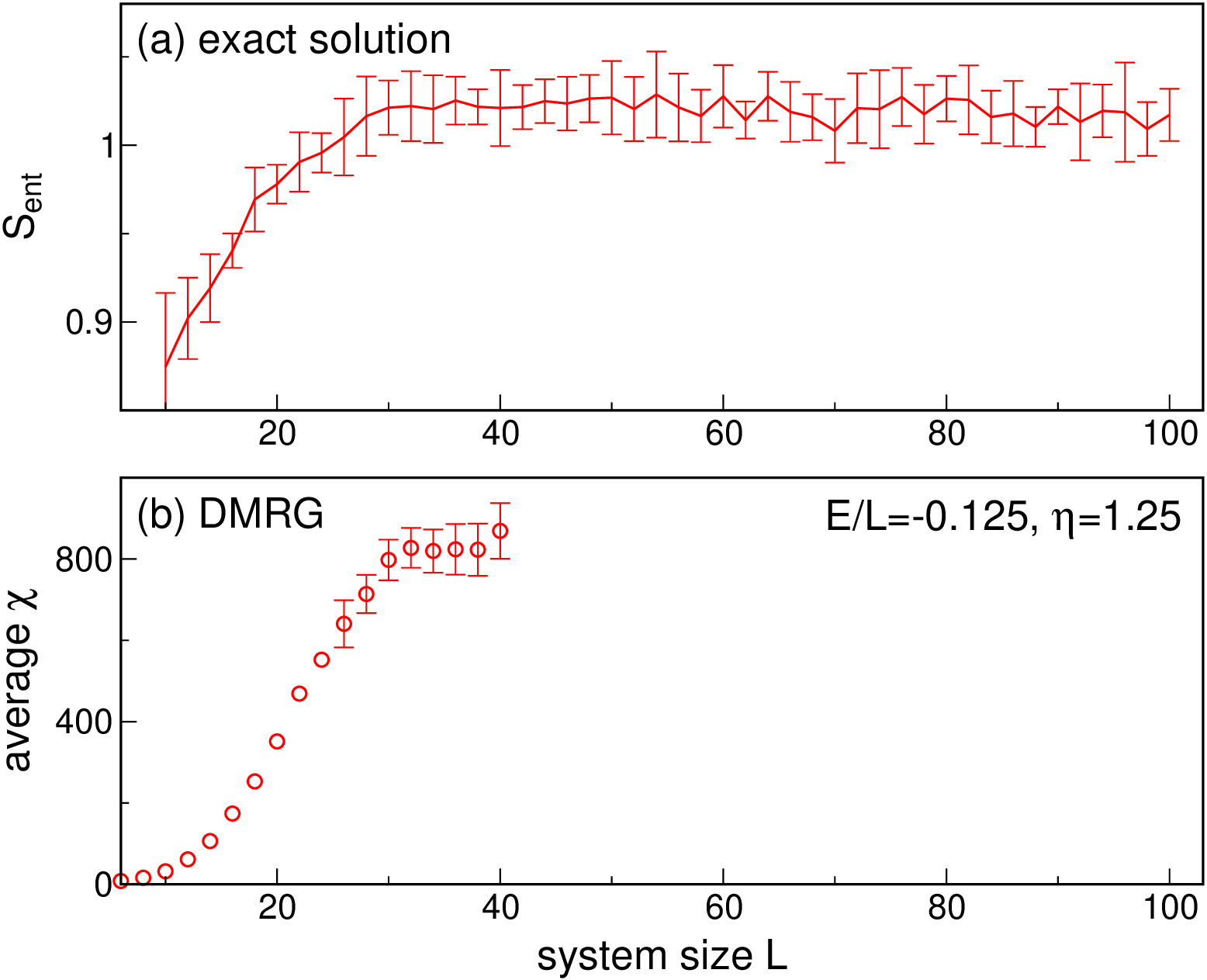}
\caption{(Color online) (a) Data from Fig.~\ref{fig:sent}(a) for $\eta=1.25$ on a magnified scale. (b) Growth of the average bond dimension during a DMRG calculation for the same parameters ($\sim100$ samples at each $L$). The maximally allowed bond dimension is $\chi\sim1700$.  }
\label{fig:test3}
\end{figure}

\subsection{Quantitative comparisons \& \\ Scaling of the entanglement at $\Delta=0$}

We now re-visit the issues raised in Sec.~\ref{sec:dmrg} and ask: Can we provide further evidence that the excited states determined by our algorithm are generic? We first calculate the system-size dependence of the bipartite fluctuations of the magnetization,
\begin{equation}
F = \langle (S^z_A)^2\rangle -\langle S^z_A\rangle^2\,,~S^z_A = \sum_{l=1}^{L/2} S^z_l~.
\end{equation}
and compare the DMRG data with the exact solution introduced in Sec.~\ref{sec:exact}. We start out with $F$ instead of the entanglement $S$ since it is expected to feature similar scaling behavior \cite{edluitz} but exhibits smaller fluctuations and is hence better suited for a quantitative comparison. Results are shown in Figure \ref{fig:test} for a fixed $E/L=-0.125$ and various $\eta$, indicating that our method can indeed be used to study the generic properties of states at a given energy density.

We now switch to the entanglement itself. To the best of our knowledge, no exact data for the scaling of $S$ in excited states at $\Delta=0$ has been published so far \cite{sentfootnote}. Hence, it is instructive to first discuss the exact results, which we show in Figure \ref{fig:sent}(a) for a fixed energy $E/L=-0.125$. In a clean system ($\eta=0$), the volume law $S(L)\sim L$ holds. Naively, one would expect that upon switching on disorder one observes a crossover into an area law $S(L)\sim\tn{const.}$ on a scale set by the localization length. However, our data indicates that in between those two limits a large regime exists where $S$ grows logarithmically. This is illustrated by the inset to Fig.~\ref{fig:sent}(a).

We now investigate the entanglement entropy in the non-interacting case using our excited-state DMRG approach. As mentioned above, $S$ exhibits fluctuations which are much larger than those of $F$; hence, significantly larger sample sizes are needed, and we restrict the comparison to a single parameter set at $\eta=5$. Figure \ref{fig:sent}(b) illustrates that the DMRG data is in decent agreement with the exact solution (the curve with $\Delta>0$ will be discussed in Sec.~\ref{sec:mblresults}).

\begin{figure}[t]
\includegraphics[width=0.95\linewidth,clip]{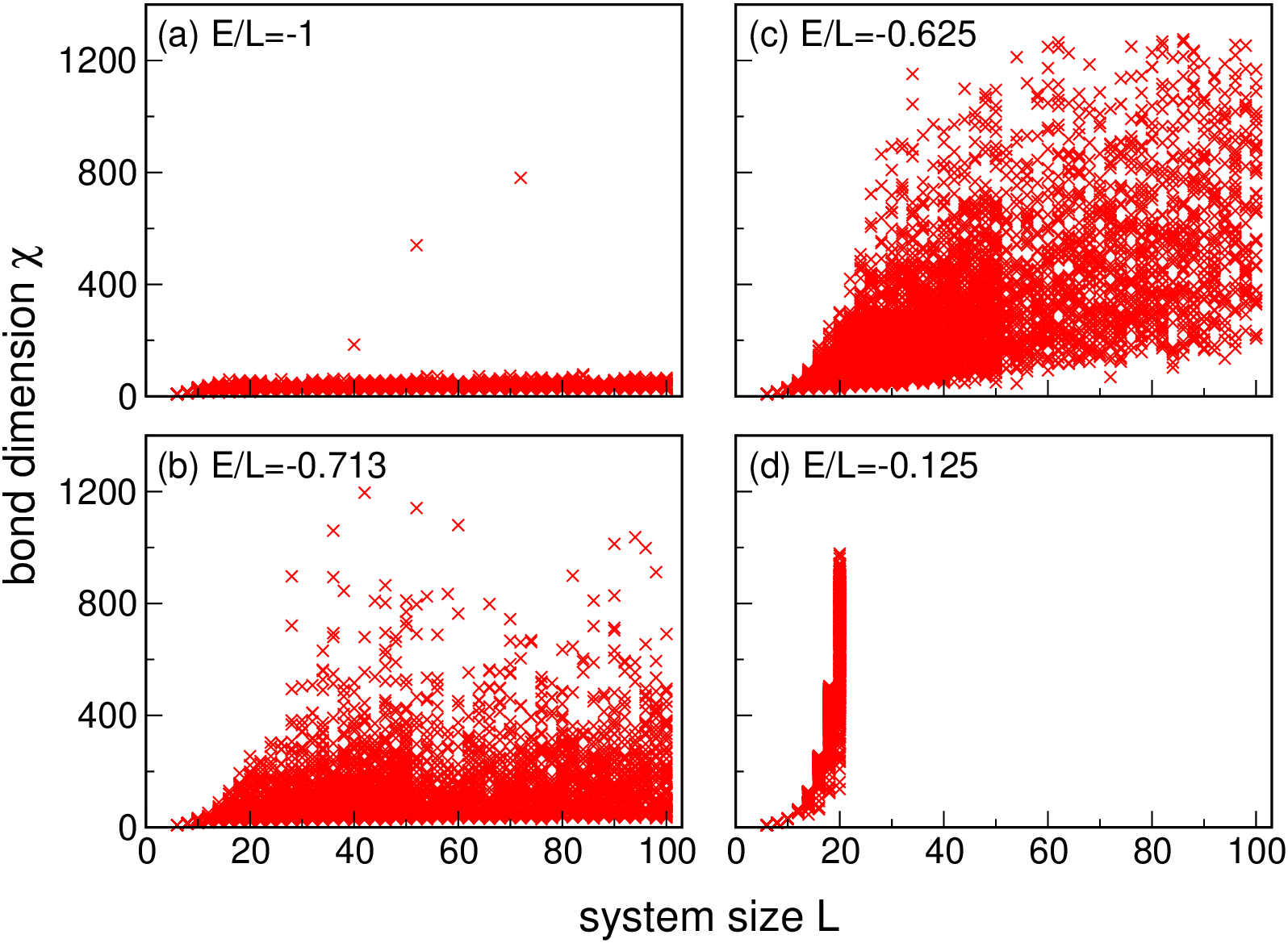}
\caption{(Color online) The same as in Fig.~\ref{fig:scatter1} but for an isotropic XXZ chain ($\Delta=1$), fixed disorder $\eta=2.5$ but various energies $E/L$. Each panel shows data for $O(5000)$ configurations. }
\label{fig:scatter2}
\end{figure}

Instead of calculating the entanglement entropy directly, one can study the scaling of the average bond dimension $\chi$ which exhibits smaller fluctuations and whose behavior is thus easier to resolve numerically. Both quantities are qualitatively related via $S\sim\log(\chi)$. Results are shown in Fig.~\ref{fig:test3}(b); the parameters coincide with those of Fig.~\ref{fig:test3}(a), which displays the corresponding exact result for $S$. An initial increase of both $S$ and $\chi$ is followed by a saturation on the same scale; the qualitative behavior is the same.

The scaling of the bipartite fluctuation shown in Fig.~\ref{fig:test} and the scaling of the entanglement (or bond dimension) shown in Figs.~\ref{fig:sent}(b) and \ref{fig:test3} provide further evidence that our method allows to determine the generic behavior of eigenstates at the targeted energy. Most importantly, the fact that our algorithm reproduces the exact result for $S(L)$ up to $L=100$ [see Fig.~\ref{fig:sent}(b)] is an indication that it does not yield a superposition of eigenstates with an artificially-enlarged entanglement. We have also compared the results of our algorithm against exact diagonalization data for small $L$, further supporting this statement.

\begin{figure*}[t]
\includegraphics[width=0.48\linewidth,clip]{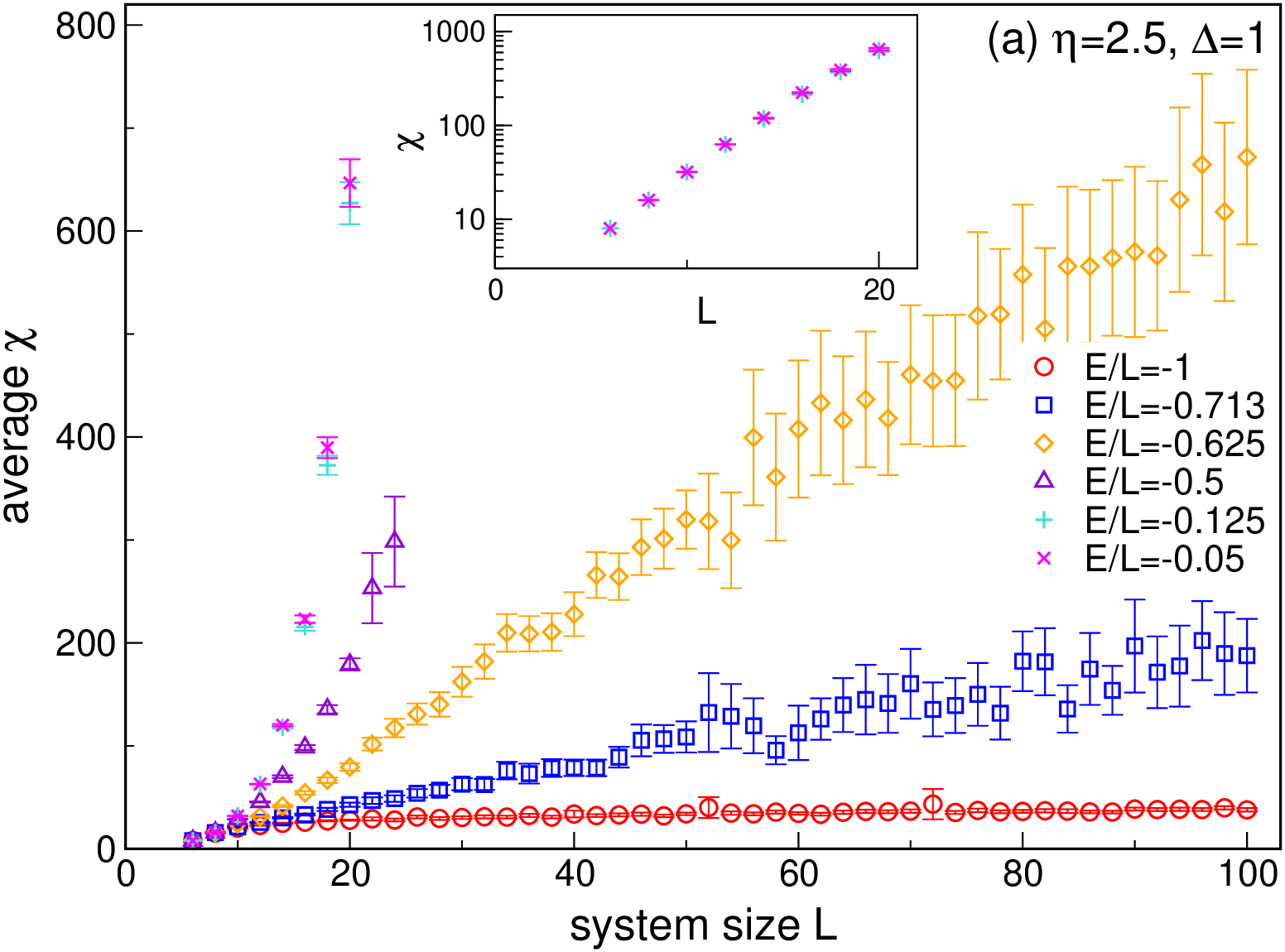}\hspace*{0.02\linewidth}
\includegraphics[width=0.48\linewidth,clip]{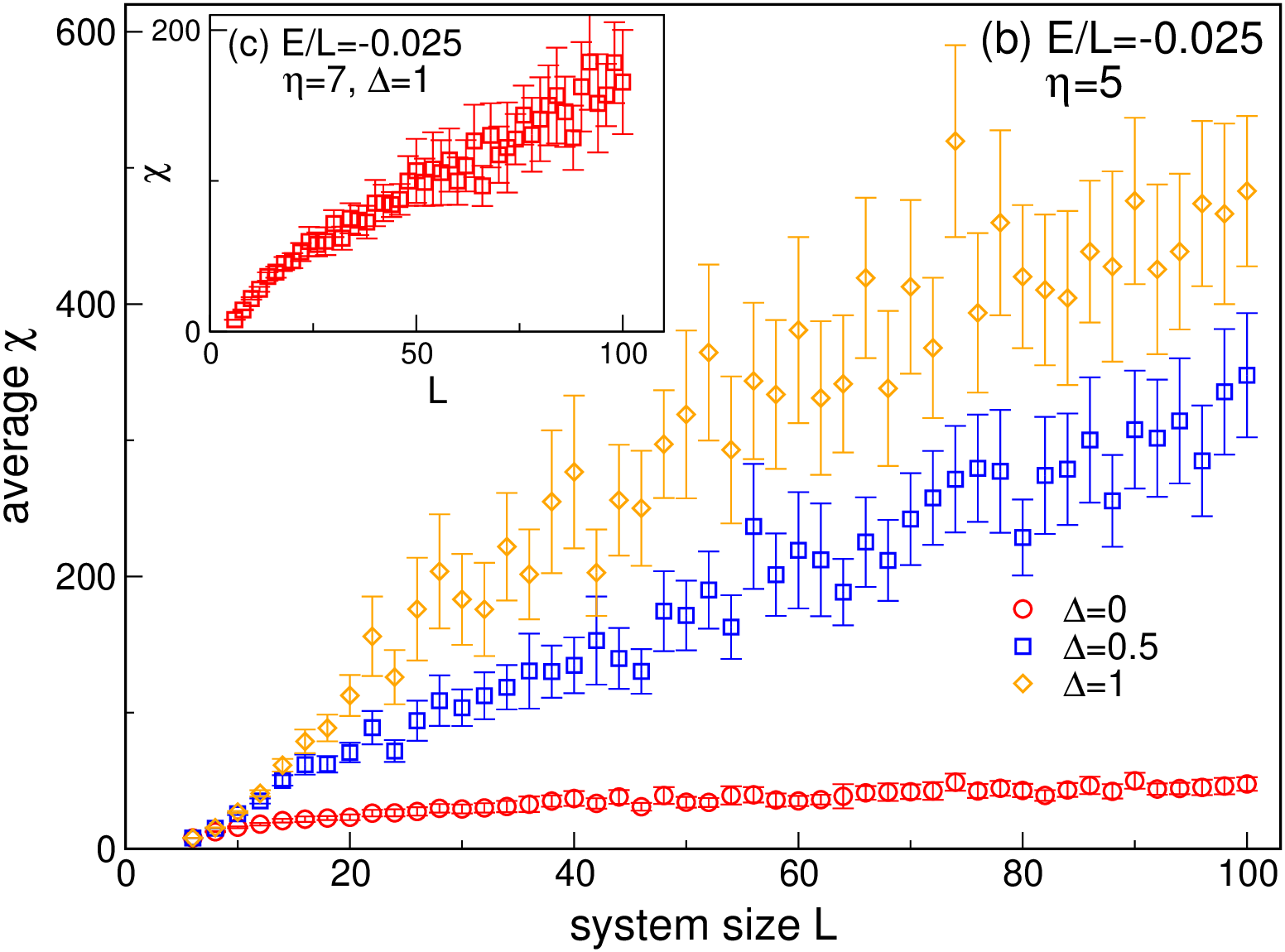}
\caption{(Color online) Scaling of the average bond dimension. The metallic phase is characterized by $\chi\sim e^{L}$; the entanglement features a volume law $S(L) \sim L$. In the many-body localized regime, our data suggests $\chi\sim L$ and hence a logarithmic scaling of the entanglement on this length scale. (a) Fixed, intermediate-strength disorder $\eta=2.5$ and various excited-state energies $E/L$ for an isotropic XXZ chain ($\Delta=1$). The inset shows the same data for the largest energies on a log-linear scale. The results are consistent with the existence of a mobility edge. (b,c) Fixed $E/L=-0.025$, various $\Delta$, and larger values of $\eta=5$ and $\eta=7$ (inset) deeper in the MBL phase. The system is localized.  }
\label{fig:bond}
\end{figure*}

\section{Many-body localized regime}
\label{sec:mblresults}

\subsection{Raw data}

We now use our DMRG algorithm to investigate the many-body localization physics of excited states. As mentioned above, similar studies (using more elaborate algorithms) appeared shortly before the completion of our paper \cite{mbldmrg1,mbldmrg2,mbldmrg3}. Our results were obtained using long-term, large-scale numerics (up to bond dimensions of $\chi\sim1700$) and hence complement and extend the data presented in Refs.~\cite{mbldmrg1,mbldmrg2,mbldmrg3}.

The isotropic XXZ chain ($\Delta=1$) was recently diagonalized exactly for up to $L=22$ sites \cite{edluitz}, indicating that all states in the spectrum are localized if $\eta\gtrsim3.5$. The data for weaker disorder is consistent with a coexistence of metallic states at high energies and localized states at the lower end of the spectrum. However, it was later conjectured that these results are plagued by severe finite-size effects, and the existence of a mobility edge was disputed \cite{mbledge}.

Fig.~\ref{fig:scatter2} shows a scatter plot of the bond dimensions occurring during the calculation of excited states at $\eta=2.5$ for various energy densities. As a reminder, we employ a two-site DMRG algorithm using a fixed discarded weight; hence, the bond dimension automatically increases to encode the amount of entanglement present in the targeted state. One can see that while the vast majority of states -- each panel shows $O(5000)$ configurations in total -- exhibits a bond dimension centered around a certain window, rare states with high entanglement exist at each $\eta$ and $E$.

Finally, a comment about the energy scale associated with the parameters of Fig.~\ref{fig:scatter2} is in order. For each individual disorder configuration, the lower and upper edges of the spectrum can be determined using ground state DMRG for $\pm H$. While states with $E/L=-0.125$ correspond to high-energy states near the center of the spectrum, those with $E/L=-1$ are of much lower energy on a scale set by the total bandwidth but still significantly away from the ground state energy (which is located at $E/L\approx-1.5$) and still belong to a dense part of the spectrum with exponentially small level spacings.

\subsection{Scaling of the entanglement; mobility edge}
\label{sec:mblresults.sent}

In Fig.~\ref{fig:sent}(b) we show for one example that the entanglement in the many-body localized phase exhibits a large intermediate regime of logarithmic growth, $S(L)\sim\log(L)$. This is analogous to the non-interacting case. As mentioned above, very large sample sizes are necessary to average out the oscillations in $S$ and to observe the logarithmic behavior. This is numerically highly demanding. Hence, we will now shift our discussion to the average bond dimension $\chi$ whose behavior can be resolved using fewer disorder configurations. Both quantities are qualitatively related via $S\sim\log\chi$.

In Fig.~\ref{fig:bond}(a), we show how the average bond dimension at $\Delta=1$ and intermediate disorder $\eta=2.5$ scales with the system size. At low energy densities (large $-E/L$), $\chi$ grows linearly with $L$, which again implies $S(L)\sim\log(L)$. As the energy increases, we observe a sharp crossover to an exponentially-growing $\chi$ [see the inset to Fig.~\ref{fig:bond}(a)] and hence a volume-law scaling of $S$ associated with a metallic phase. Our results are thus consistent with the existence of a mobility edge at $\Delta=1$ and $\eta=2.5$ in agreement with the results of Ref.~\cite{edluitz}. The data for larger disorder is consistent with full many-body localization -- states at all energies do not fulfill a volume law up to $L=100$ sites. This is illustrated in Fig.~\ref{fig:bond}(b,c) for larger values of $\eta=5$ and $\eta=7$ lying deeper in the MBL phase. 

While in the metallic regime the system sizes that can be tackled by our numerics are similar to those accessible by exact diagonalization, much larger $L$ can be treated by the excited state DMRG in the MBL phase. In fact, our data suggests that for both $\eta=2.5$ and $\eta=5$ one still observes transient behavior at $L\sim20$. E.g., the entanglement in mid-spectrum states at $\eta=5$ grows logarithmically even up to $L\sim100$, and the area law regime of constant $S$ has not yet been reached. This suggests complications in finite-size scaling for small $L$; employing large systems -- even beyond what is accessible by our method -- seems essential in order to determine properties of the MBL phase. Another interesting and unresolved question pertains to what scales govern the crossover between the regimes of $S(L)\sim L$ and $S(L)\sim\log(L)$ and the regimes of $S(L)\sim\log(L)$ and $S(L)\sim\tn{const.}$.

\begin{figure}[t]
\includegraphics[width=0.95\linewidth,clip]{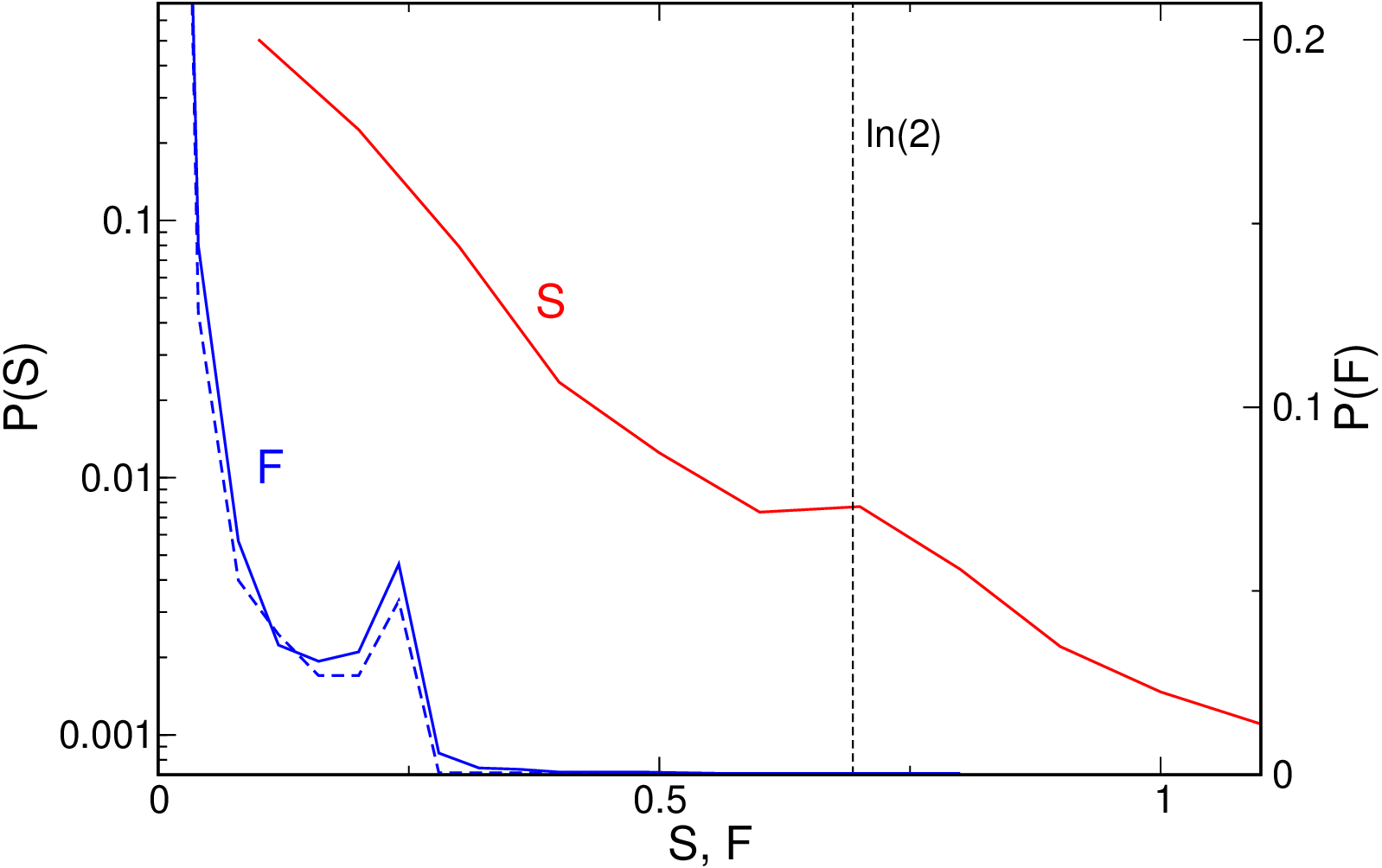}
\caption{(Color online) Probability distributions $P(S)$ and $P(F)$ for the entanglement and the bi-partite spin fluctuations of a system in the MBL phase with $\eta=7$, $\Delta=1$, $E/L=-0.025$, and $L=10$ (solid) or $L=20$ (dashed). The data was obtained using $O(3000)$ samples. }
\label{fig:hist}
\end{figure}

For reasons of completeness, we finally show the probablity distribution $P(S)$ of the entanglement in Fig.~\ref{fig:hist}; it exhibits a peak at $ln(2)$ in agreement with the results of Refs.~\cite{mbldmrg2,mbldmrg3}. Similarly, the probability distribution $P(F)$ of the bi-partite spin fluctuations exhibits a peak at $F=1/4$.

\subsection{Eigenstate thermalization hypothesis}
\label{sec:mblresults.eth}

\begin{figure}[b]
\includegraphics[width=0.95\linewidth,clip]{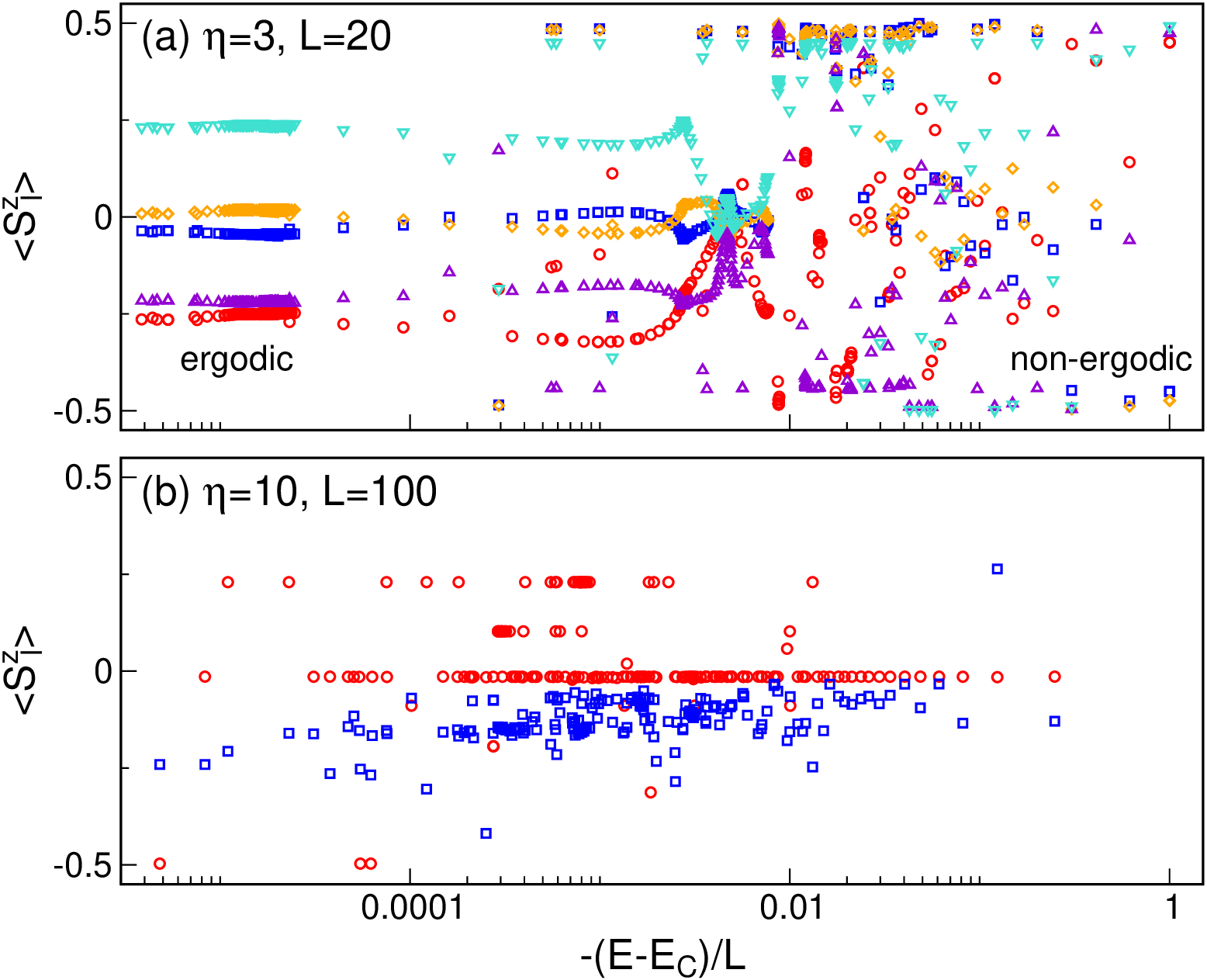}
\caption{(Color online) Probing the eigenstate thermalization hypothesis via the expectation value of the local magnetization $\langle S^z_l\rangle$ for $\Delta=1$ and at various energies. (a) Single configuration drawn at $\eta=3$ for a chain of $L=20$ sites (five different positions $l\approx L/2$ are shown; the disorder configuration is the same at each $E$). One observes a crossover between thermal and non-ergodic behavior as the energy decreases. (b) Two single configuration drawn at $\eta=10$ for a large chain with $L=100$ (only one position $l=L/2$ is shown for each). States at all energies are localized and non-ergodic.  }
\label{fig:eth}
\end{figure}

Another key feature of localized systems is non-ergodic behavior. In a static setup, this can be investigated by checking the eigenstate thermalization hypothesis (ETH), which conjectures that the expectation values of local observables coincide for all states which are close in energy. Fig.~\ref{fig:eth} displays the energy-dependence of the local magnetization $\langle S^z_l\rangle$ for \textit{single disorder configurations} (not averages). Energies are shifted with respect to the center $E_\tn{C}$ of the spectrum of the corresponding individual configuration, which we determine by targeting the highest state of $H$ using conventional ground state DMRG for $-H$.

Fig.~\ref{fig:eth}(a) shows data for $\Delta=1$ and $\eta=3$ for a system of size $L=20$; we expect a coexistence of metallic and localized states. Indeed, the ETH is violated at low energies (large negative $E$) but holds at high energies. Our DMRG algorithm can now be used to tackle much larger lattices in the MBL phase to demonstrate that states are non-ergodic at all energies. Results are shown in Fig.~\ref{fig:eth}(b) for $L=100$; a clear violation of the ETH can be observed.

\section{Outlook}
\label{sec:outlook}

We have studied the behavior of excited states in large many-body localized systems using a modified density matrix renormalization group algorithm. The method was benchmarked against exact results constructed in the non-interacting limit. In the MBL phase, there is a large domain of logarithmic entanglement growth separating the regimes where volume and area laws hold. This illustrates the need to tackle this problem by methods beyond exact diagonalization. Our results are consistent with the existence of a mobility edge.

Expressing lowly-entangled excited states of large localized systems in terms of a matrix product state was a long-standing open problem. One can envision a plethora of future applications for the algorithm presented in this work as well as for the closely-related (more elaborate) ideas discussed in the preprints of Refs.~\cite{mbldmrg1,mbldmrg2,mbldmrg3}. Studying the dynamics of perturbations in excited states is one possible avenue.

\emph{Acknowledgments} --- We thank David Huse, Dong-Ning Sheng, Frank Pollmann, and Romain Vasseur for useful comments. DK acknowledges support by the Deutsche Forschungsgemeinschaft via RTG 1995. CK acknowledges support by the Emmy Noether program of the Deutsche Forschungsgemeinschaft (KA 3360/2-1).

\begin{thebibliography}{*}
\bibitem{sentfootnote}
The scaling of the entanglement in the non-interacting limit was also studied in Ref.~\cite{sent3}. For small disorder, the authors observe logarithmic behavior which is a remainder of the area law $S(L)\sim \frac{1}{3}\log(L)$ associated with the clean case. This is \textit{not} the logarithm we observe, which at low energies occurs on a much smaller scale and is not resolved in Ref.~\cite{sent3} where only the ground state is studied.
\end{thebibliography}
\end{document}